\pdfoutput=1
\documentclass{acmsmall}
\makeatletter
\def\endbottomstuff{
\end@float
}
\def\firstfoot{}
\runningfoot{}
\makeatother

\usepackage[ruled]{algorithm2e}

\newcounter{enum2}

\newcommand{\calC}{\mathcal{C}}
\newcommand{\calD}{\mathcal{D}}

\def\argmax{\mathop{\rm arg\,max}}

\begin{document}

\markboth{M. Momma et al.}{Influence Analysis in the Blogosphere}

\title{Influence Analysis in the Blogosphere}
\author{MICHINARI MOMMA
\affil{GREE Corp.}
YUN CHI
\affil{NEC Laboratories America}
YUANQING LIN
\affil{NEC Laboratories America}
SHENGHUO ZHU
\affil{NEC Laboratories America}
TIANBAO YANG
\affil{Michigan State University}}

\begin{abstract}
  In this paper we analyze influence in the blogosphere.  Recently,
  influence analysis has become an increasingly important research
  topic, as online communities, such as social networks and e-commerce
  sites, playing a more and more significant role in our daily life.
  However, so far few studies have succeeded in extracting influence
  from online communities in a satisfactory way.  One of the
  challenges that limited previous researches is that it is difficult
  to capture user behaviors.  Consequently, the influence among users
  could only be inferred in an indirect and heuristic way, which is
  inaccurate and noise-prone.  In this study, we conduct an extensive
  investigation in regard to influence among bloggers at a Japanese
  blog web site, BIGLOBE.  By processing the log files of the web
  servers, we are able to accurately extract the activities of BIGLOBE
  members in terms of writing their blog posts and reading other
  member's posts.  Based on these activities, we propose a principled
  framework to detect influence among the members with high
  confidence level.  From the extracted influence, we conduct in-depth
  analysis on how influence varies over different topics and how
  influence varies over different members.  We also show the
  potentials of leveraging the extracted influence to make
  personalized recommendation in BIGLOBE.  To our best knowledge, this
  is one of the first studies that capture and analyze influence in
  the blogosphere in such a large scale.
\end{abstract}




\begin{bottomstuff}
The work of the first author was done when he was at NEC Labs.

Author's addresses: Y. Chi {and} Y. Lin {and} S. Zhu, NEC Laboratories
America, Cupertino, CA 95014 USA; T. Yang, Michigan State University,
East Lansing, MI, USA.
\end{bottomstuff}

\maketitle

\section{Introduction}\label{sec:introduction}
Influence analysis is a very important research topic in social
science and is becoming more and more important in online communities
as online social networks, such as Facebook and Twitter, and online
e-commerce sites, such as Amazon and Netflix, playing increasingly
important roles in people's daily life.  In these online communities,
influence is ubiquitous: in a social network, the activities and
interests of a user (e.g., what a blogger reads or writes about) are
usually heavily affected by that of his or her friends in the network;
in an online e-commerce site, the opinions of an authoritative
reviewer can significantly sway the purchase decisions of many
customers.  Analyzing such influence, in addition to serving
scientific research purposes, also has practical importance in various
areas.  For example, it may offer accurate opinion survey for
politicians or play a key role in product promotion and damage control
for businesses.

Social influence describes the phenomenon by which the behavior of an
individual is directly or indirectly affected by the thoughts,
feelings, and actions of others in a
population~\cite{kraut98,song07,cialdini08}.  As can be seen, there
are two important components in social influence. The first component
is the behavior or actions of an individual, and the second component
is that these actions should be a consequence of \textit{being
  affected} by other people.  These two components rely on a
\textit{causal} effect between the actions of an individual and that
of other people that influence the individual.  To detect such a
causal effect in the influence is a very challenging problem.  Most
existing approaches adopt certain heuristics for detecting influence,
e.g., by considering the temporal order of actions (user Alice is
influenced by user Bob if Alice uses the same keywords
\cite{EytanAdar} or the same tags \cite{influenceAndCorrelationInSN}
\textit{after} Bob has done so).

These heuristics, however, failed to distinguish between the effect of
\textit{causality} and that of \textit{correlation}.  We use the
blogosphere to illustrate this point.  Assume blogger Alice and
blogger Bob each writes a post on the topic of healthcare reform and
Alice's post dated later than that of Bob.  If such actions are
observed, can we draw a conclusion that blogger Alice has been
influenced by blogger Bob on the topic of healthcare reform?  Such a
conclusion is obviously flawed because there may exist other reasons,
other than Alice being influenced by Bob, for Alice and Bob to write
similar posts---maybe there was a news event about healthcare reform
that triggers both Alice's and Bob's posts.  In other word, we may
claim Alice and Bob are \textit{correlated}, but should not establish
\textit{causal} relationship no matter which post is written first.
Separating causality from correlation is a notoriously difficult
problem \cite{influenceAndCorrelationInSN}.  The difficulty is partly
due to that in real applications, the ground truth is usually not
available in all but a few cases.\footnote{The paper citation network
  is such a rare case because the author of a paper usually explicitly
  declares the source of influence for the paper in its reference. Of
  course, it is not totally noise-free, due to the existence of
  bias~\cite{howCitation}. } Without the ground truth, no one can
claim they separate causality from correlation \textit{with certain}.

Facing such a challenge, in this paper we propose a method to
determine, with high level of confidence, the influence among bloggers
in the blogosphere.  More specifically, in this work, we investigate
the influence in a close-world blogosphere, the BIGLOBE blog community
Webryblog\footnote{\texttt{www.biglobe.ne.jp} and
  \texttt{webryblog.biglobe.ne.jp}.}. 

BIGLOBE is one of the leading Internet service providers in Japan and
it provides various portal services including a blog service called
Webryblog to its members.  From the web server log files, we are able
to capture the activities of BIGLOBE members.  In this work, we mainly
focus on two types of actions among BIGLOBE members: writing posts and
reading other member's posts.  By studying these actions, we propose a
framework to identify, with high confidence level, influence among
members. Employing real actions to identify influence is a major
contribution of this paper.  From the identified influence, we are
able to conduct in-depth analysis on how members influence each other
in BIGLOBE.  To our best knowledge, this is the first analysis of
influence, where influence is relatively accurately identified, in
such a large scale.

After obtaining the influence among its members, we are able to answer
various questions about influence in Webryblog.  In this paper we
focus on two questions: ``Are there different influential bloggers on
different topics?'' and ``Are there different influential bloggers for
different members, even on the same topic?''.  Intuitively, the
answers to both the questions should be \textit{yes}.  For the first
question, as an anecdotal proof, if we look at the top-100 popular
blog list at
Techorati\footnote{\texttt{http://technorati.com/blogs/top100/}.},
which is an authoritative blog ranking site, we can see that most of
the top popular (influential) blogs only focus on a special domain
(politics, technology, celebrity gossip, etc.).  For the second
question, we again use the previous example of healthcare reform: who
are the most influential bloggers to Alice on the issue of healthcare
reform probably, given Alice has Democratic leanings.  To verify the
above intuitions, in this paper we design several tests by leveraging
some techniques we recently developed for social network
analysis. From the results of these tests, we are able to provide
affirmative answers to these two questions by using certain
quantitative measures.

The rest of the paper is organized as follows.  In
Section~\ref{sec:related}, we give a survey on related work.  In
Section~\ref{sec:datastudy}, we provide a detailed description of the
blog data set that we use.  In Section~\ref{sec:influence}, we propose
a method to detect influence from the user access log.  In
Section~\ref{sec:topic}, we investigate topic-specific influence.  In
Section~\ref{sec:experiment}, we investigate member-specific influence
and apply it to the application of blogger recommendation.  Finally,
in Section~\ref{sec:conclusion}, we conclude and give future
directions.


\section{Related work}\label{sec:related}
As mentioned in Section \ref{sec:introduction}, the isolation of
influence from other sources of correlation is known as a very
challenging issue.  Anagnostopoulos et
al.~\cite{influenceAndCorrelationInSN} addressed this issue by
proposing some statistical tests for isolating influence from social
correlations and applied to a large data set of 340K users and 2.8M
edges.
Traditionally, studying similarity between people in a social network
has been a central research focus. Gruhl et
al~\cite{informationDiffusion}. analyzed information propagation in
the blogosphere by tracking topics in blog posts. Kumar et
al.~\cite{onTheBurstyEvolution} used hyperlinks to form a blog-graph
and studied how the blog-graph grows over time. Adar et
al.~\cite{EytanAdar} introduced implicit link, or inferred link, to
address the issue of sparsity of explicit URL links for the purpose of
stable inference of information flow. To infer the implicit link, they
used some similarity measures such as explicit URL link patterns and
timing of URL link generation. They built inferred links of several
thousand links from 1000 blogs. Note this approach has to rely on URL
links to identify similarity between bloggers, and so it relied on
$indirect$ inference of implicit link.

Using instant messaging as a link medium between people, Singla and
Richardson~\cite{yesThereIsACorrelation} studied social correlation
and identified homophily by using demographic information as well as
search queries. They showed that given a link that is defined by an
instant messaging event, the probability of having the same value of
user profile, including search queries or demographic attributes, is
higher than the population average.  In~\cite{song07}, Song et
al. proposed an information flow models where the influence is
indirectly inferred by the time of adoption, e.g., innovators, early
adopters, laggards, and so on.

In \cite{analyzingPatternsOfUser}, Guo et al. studied user's posting
behavior in knowledge sharing forums in detail. In online forums, Shi
et al.~\cite{userGroupingBehavior} reported the probability of joining
a community in terms of community features, such as size, links formed
by reply-friends, and ratings of top posts. Similarity between users
can explain how many common communities the users have, and it is
defined by a frequency of direct reply relations and the number of
common friends.  In~\cite{Wang11}, Wang et al. studied and modeled how
information spreads in a large enterprise throw emails.  However, such
information spreading is mainly task-driven, e.g., among emails about
a particular consulting project.  In~\cite{Romero11}, Romero et
al. studied how information diffuses differently across topic among
Twitter users.  However, the main focus of that work is to reveal
information spreading at a macroscopic level, namely over different
topics, instead of at the level of influence among specific
individuals.  In~\cite{LaFond10} La~Fond et al. proposed a
randomization test to separate social influence (causal effects) and
homophily effects (correlation).  The proposed randomization test,
however, is again at a macroscopic level where the aggregated edge
counts related to particular attributes are used to infer influence
vs. homophily.

Once influence of each user has been identified, identifying a set of
most influential people would lead to interesting applications or
services. The problem can be formulated as a set cover maximization
problem. There has been a lot of research work for solving the problem
\cite{cost-effectiveOutbreakDetection,miningKnowledge-sharingSites,miningTheNetworkValueOfCustomers,efficientInfluenceMaximization}
proposed efficient algorithms to solve the related discrete
optimization problem.  Arini et al.~\cite{turningDownTheNoise}
addressed a personalized cover maximization problem and as an
application, personal recommendation has been proposed and evaluation
has been done by human subjects.

Given a graph made up from links of influential relations among
bloggers, we can use link prediction techniques to predict future
links. As for the link prediction \textit{in general}, HITS
\cite{hits_paper} and PHITS \cite{phits} are the link analysis
counterparts of the latent semantic analysis (LSA) that are typically
used for content analysis. These methods are all based on low rank
approximation of the matrix data with various interpretations from
linear algebra and probability.  Recently, combining link analysis
with content information, for improving prediction performance, has
been paid much attention.  Cohn and Hofmann~\cite{missingLink}
proposed a factorization based method to incorporate both link and
content information. Multi-dimensional (tensor) factorization was used
by Chi et al.~\cite{yun09:_iolap} and Chen et
al.~\cite{combinationalCollaborativeFiltering} for the problem.  Some
models that are based on the Latent Dirichlet Allocation have also
been proposed
\cite{unsupervisedPredictionCitationInfluences,jointLatentTopicModels}.
Moreover, some supervised learning methods have been developed to show
promising results compared with unsupervised
models~\cite{combiningLinkAndContentForCommunityDetection,discLDA}.

\section{Data Study}\label{sec:datastudy}
Overall we have collected data for a period of a whole year, between
September 2008 and August 2009.  At Webry Blog, there are two
different types of servers: one provides editing and posting service
and the other browsing service.  From these servers we obtained two
kinds of data. One data set is a collection of blog posts, referred to
as {\it blog content}, which is obtained from the editing servers.
The other is a collection of server access logs in the same period,
referred to as {\it access log}, which is obtained from the browsing
servers.

\begin{table*}[htbp]%
\tbl{Fields in Blog Content File and Access Log File\label{tab:fields}}{%
\begin{tabular}{l}
  Fields for blog content file: \\
  {\begin{tabular} {|c|c|c|c|c|c|c|c|}
      \hline
      {\underbar{IP address}} & {\underbar{uploadTimeStamp}} &  
      userID & URL & title & blogName & body & themes \\
      \hline
\end{tabular}}\\
\vspace{0.05in}\\
Fields for access log file: \\
{\begin{tabular} {|c|c|c|c|}
\hline
{\underbar {IP address}} & {\underbar{accessTimeStamp}} &  request &
 referrer  \\
\hline
\end{tabular}}\\
\end{tabular}}
\end{table*}%

Records obtained from the different servers are combined to conduct
various analysis on influence among bloggers, which is to be described
in later sections.
For the purpose of matching the two data sets, we use the IP address
as a key field for binding them together.  Of course, use of a cookie
would be a better choice because an IP address can change over time
and may be shared among a number of people using the same sub-network,
making the mapping between an IP address and a user
many-to-many. Unfortunately, the cookie for identifying unique users
was not available to us.  However, we noticed in our preliminary study
that by controlling the time difference, {\it i.e. window size}, of
posting and browsing behaviors, the issue of shared IP address can be
minimized\footnote{We manually inspected IP addresses and blog content
  associated with them to estimate noise. We observed even if an IP
  address is shared by different bloggers, the upload time is
  typically different. Therefore, minimizing the window size will
  reduce the instances of the shared IP address.}.

\subsection{Blog Content}
The data fields for blog posts are summarized in
Table~\ref{tab:fields}.
The raw value of IP address is hidden at BIGLOBE via a one-way hash
mapping for preserving privacy.\footnote{Since the reverse operation
  is impossible, there is no way to recover the raw value from the
  data.}
 Uniqueness of the anonymized IP address remains valid so the matching
 operation is still valid as well.
 The uploadTimeStamp is the time at which a post is uploaded for the
 first time. Even if the blogger edits the post later, the timestamp
 does not change. So there is no way to tell if a post is modified
 after its first uploading. URL is the URL of the post, title is the
 title of the post, userID is identical to the domain name of the blog
 webpage of the blogger, blogName is the name of the series of blog
 posts by the blogger, body is the content of the post. Themes are the
 themes of the post and they are in a free description format, as
 opposed to a set of predefined categorical tags. In the editing site,
 the system shows popular themes from which the blogger can choose as
 the themes for their post, and this results in many bloggers using
 popular themes for their posts.

\begin{figure}[ht]
\centering
\includegraphics[width=\linewidth]{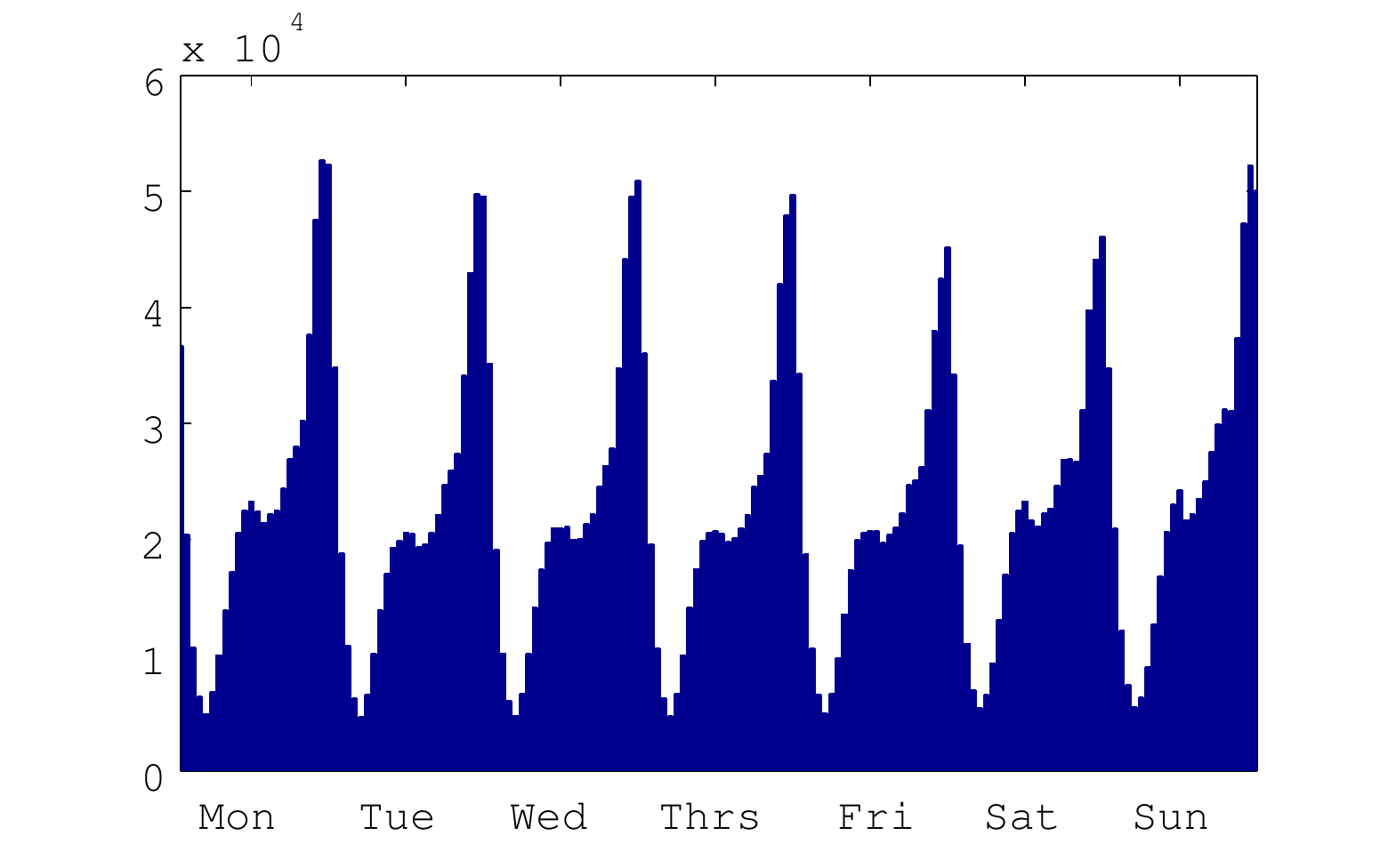}
\caption{The number of blog posts for week days (where actually the
  hourly volumes are shown).} \label{fig:blogPostWeek}
\end{figure}

\begin{table}[h]%
  \tbl{Frequent themes: column 1 is for the rank, column 2 for
    ratio, and column 3 for the theme.\label{tab:popularThemes}}{%
\begin{tabular} {|c |c| c|}
\hline
Rank & Percentage(\%) & Theme \\
\hline
1 & 10.7  & diary \\
\hline
2 & 8.00 & monologue \\
\hline
3 & 3.41 & notes \\
\hline
4 & 3.15 & everyday life \\
\hline
5 & 2.21 & photograph \\
\hline
6 & 2.10 & life \\
\hline
7 & 1.97  & music \\
\hline
8 & 1.66  & flower \\
\hline
9 & 1.55 & mumble \\
\hline
10 & 1.51  & game \\
\hline
11 & 1.47  & gourmet \\
\hline
12 & 1.46 & travel \\
\hline
13 & 1.40 & movie \\
\hline
14 & 1.32 & children \\
\hline
15 & 1.29 & news \\
\hline
\end{tabular}}
\end{table}%

During the period, the total number of blog posts was 3,870,520.  The
average number of post per day was 11,059.  Figure
\ref{fig:blogPostWeek} shows a weekly pattern of the volume of
postings. As expected, a periodical trend is noticeable. Typically,
Sundays have the highest volumes. Also, though not shown in the
figure, holidays have higher volumes, suggesting bloggers typically
write blogs on non-working days.  Qualitatively,
Figure~\ref{fig:blogPostWeek} has a very similar shape to the
corresponding blog posting frequency distribution in
\cite{analyzingPatternsOfUser}.  This posting tendency can be
explained by the popularity of diary-related blogs, since people
usually write their diaries after some social events occurred and such
events typically happen on Sundays or
holidays. Table~\ref{tab:popularThemes} shows a ranking of frequent
blog themes. Note that the top four themes are all related to diaries
and the total sum of them is more than 20\% of the total blog posts.

\begin{figure}[h]
\centering
\includegraphics[width=\linewidth]{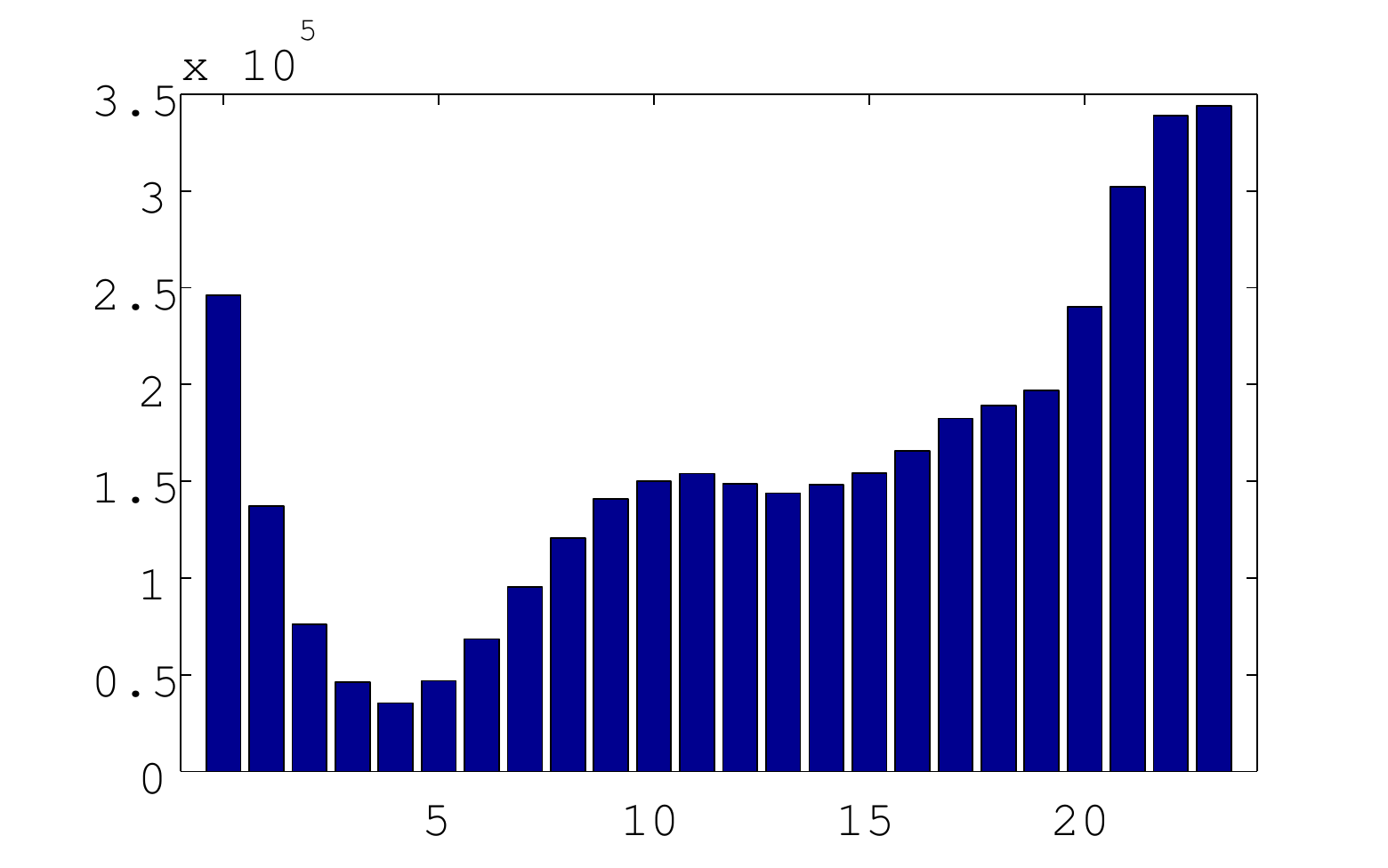}
\caption{The number of blog posts for each
  hour.} \label{fig:blogPostHour}
\end{figure}

Figure \ref{fig:blogPostHour} shows how the number of blog posts
changes within a day. Not surprisingly, bloggers write blogs late at
night and become less active during the daytime. Again, the shape is
very similar to its counterpart in \cite{analyzingPatternsOfUser}.

\begin{figure}[h]
\centering
\includegraphics[width=\linewidth]{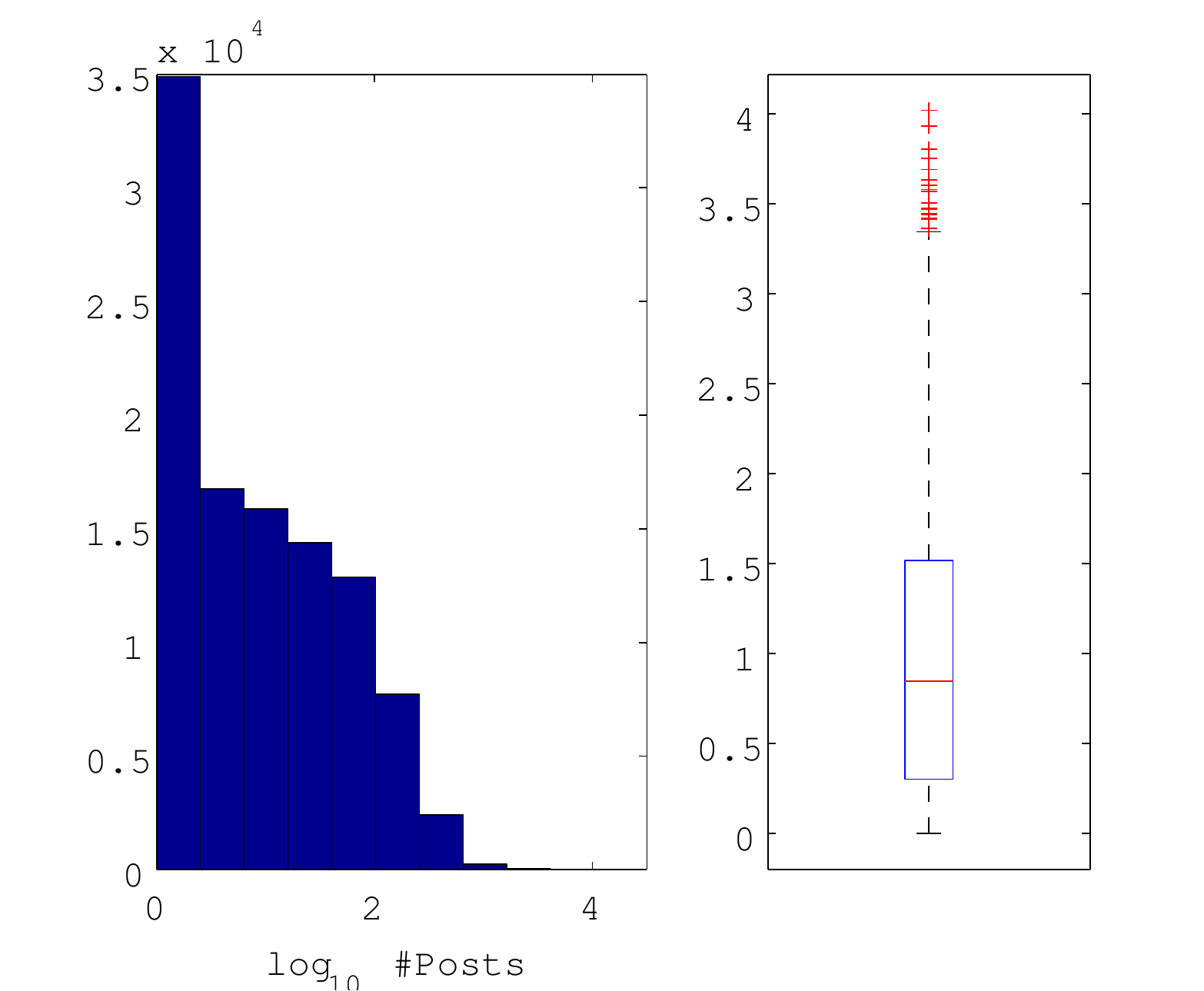}
\caption{{\it left:} Distribution of the number of posts per
 blogger. The number of post is on the common log scale.
{\it right:} Box plot of the distribution. }\label{fig:distNumPosts}
\end{figure}

Figure \ref{fig:distNumPosts} shows the histogram and box plot of the
number of posts per blogger. The distribution is highly skewed. The
mean number of posts is 37.5 while the median is 7. The skewness is
due to the high volume of frequent bloggers who sit in the long tail
of the distribution.  Also, as revealed in the box plot, there are
some bloggers who wrote extremely large number of posts.

\subsection{Access Log}\label{sec:accesslog}
The format of the access log files follows the Apache combined
format. Table \ref{tab:fields} summarizes the fields used in our
data analysis.  IP address is the IP address associated with the
access, accessTimeStamp is the timestamp of the access, request
contains processing requests, and referrer is the referrer of the
access, which is only used for removing auto-generated accesses later.
In particular, to match a blog content and a record in the access log,
IP address, uploadTimeStamp, and accessTimeStamp are used as a
composite {\it soft} key.

The access logs contain all information of server accesses, which
makes the number of the records massive. We remove unnecessary records
to make the data processing that follows more efficient.
Since the access log is only used for analyzing bloggers' access
pattern, accesses from non-bloggers are ignored.
Additionally, accesses due to RSS feeds and robots are removed by
using the referrer field. Further, we identify some IP addresses that
are associated with anomalously large numbers of access logs. Since
such accesses seem to be generated by automated measures, we remove
records from such IP addresses as well.
The following summarizes the criteria of removing records:
\begin{itemize}
\item records associated with IP address that does not appear in blog content during the data collection period
\item records associated with access to their own articles
\item records associated with access to index.html (contents are not available)
\item records associated with access to non-html (images, etc)
\item records associated with more than 12 hours before and after posting
\end{itemize}
Note the last criterion is set due to the following conflicting
considerations: (1) we want to reduce noise, by minimizing the time
window, because of the issue of duplicated IP address in identifying
bloggers; (2) however, we do not want to restrict the time window too
close to the upload time. In other words, we want to minimize the
window size to remove noise but at the same time we want to maximize
it to study {\it general} behavior of bloggers. We determine the
window size 12 hours as appropriate for the purpose of the study by
inspecting and balancing the above conflicting aspects.


\begin{figure}[ht]
\centering
\includegraphics[width=\linewidth]{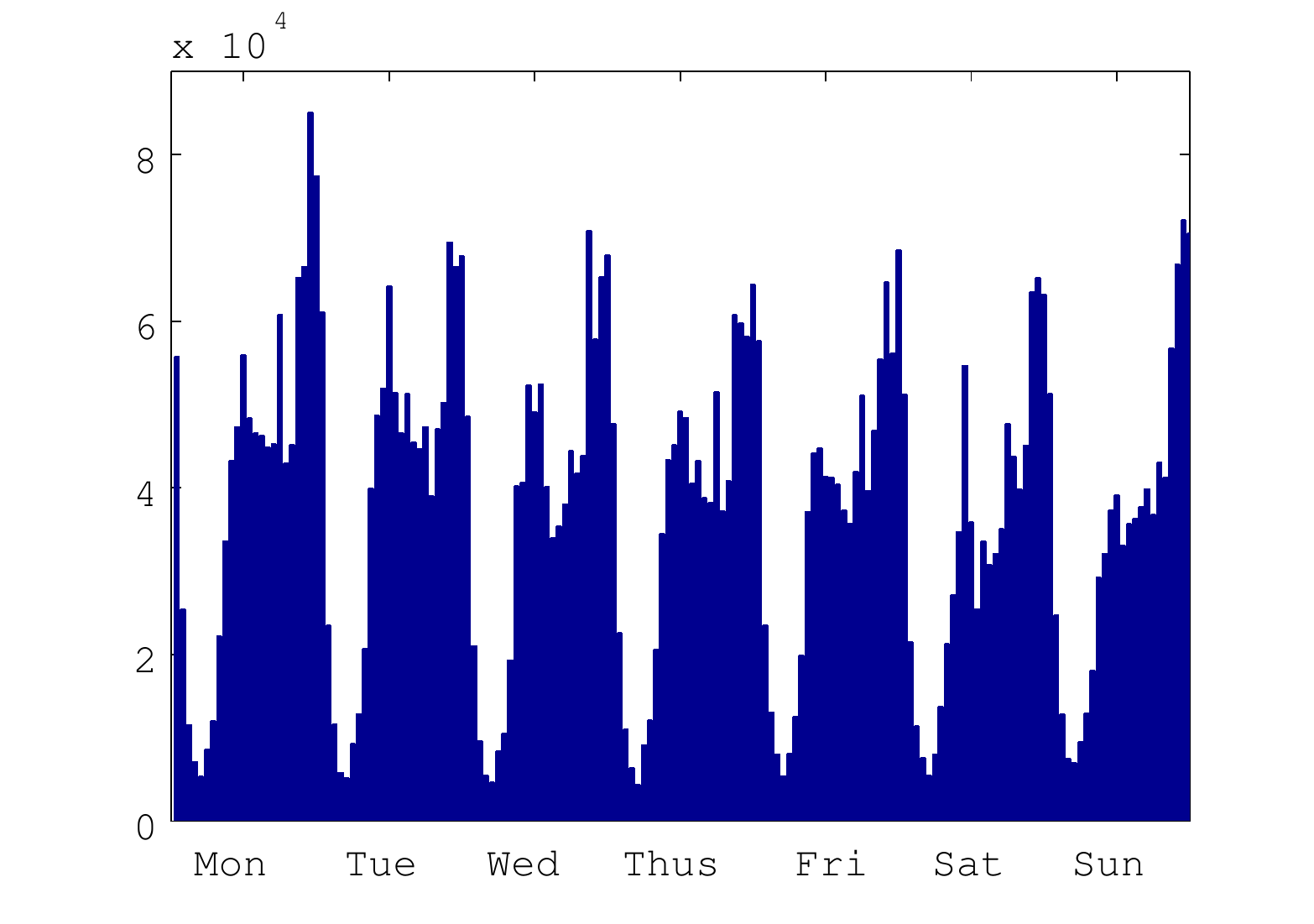}
\caption{The number of clicks to other blogger's posts for
 each day of the week. }\label{fig:accessCountsWeek}
\end{figure}

\begin{figure}[ht]
\centering
\includegraphics[width=\linewidth]{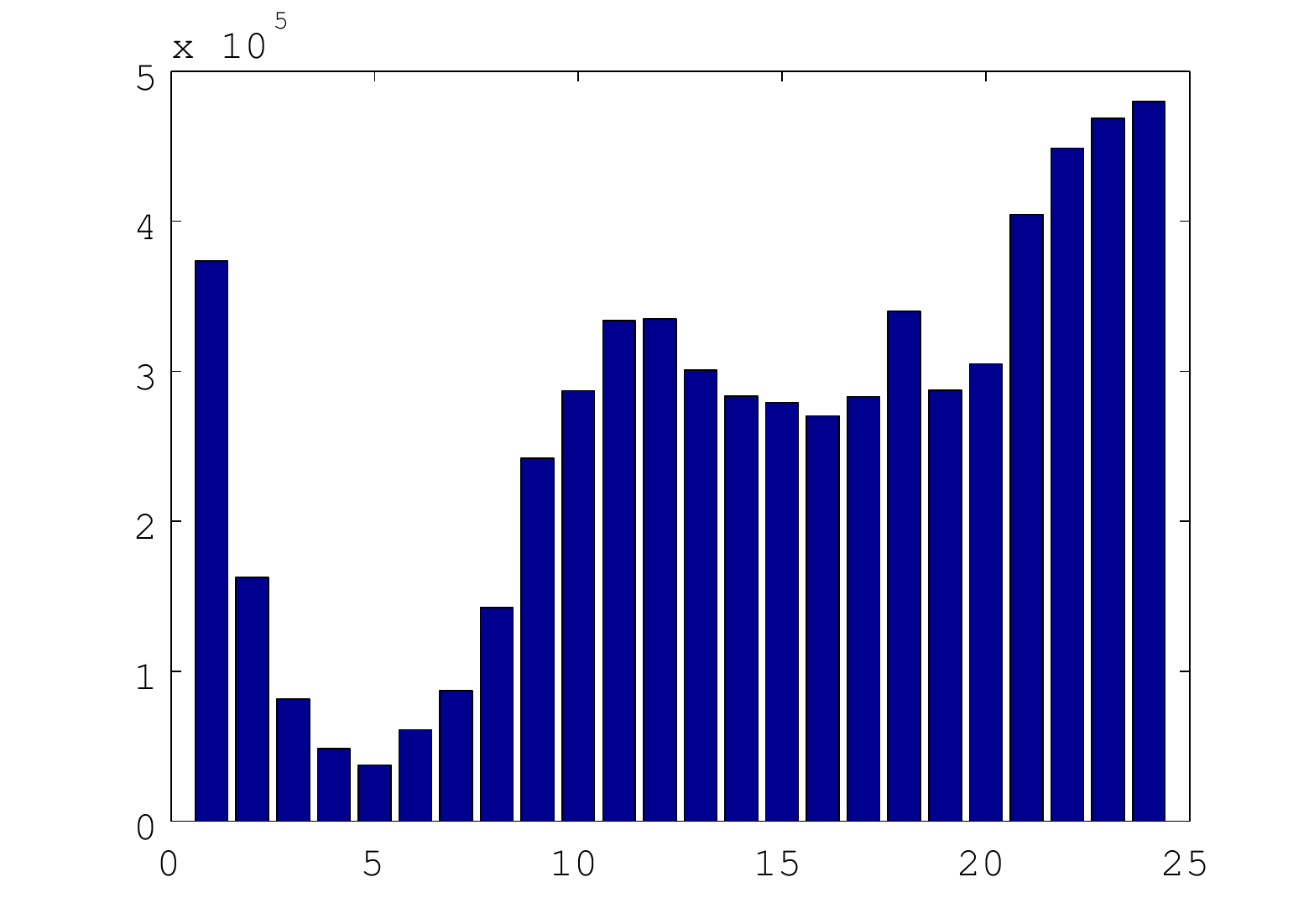}\\
\caption{The number of clicks to other blogger's posts for each hour
  of the day. }\label{fig:accessCountsDay}
\end{figure}

Figure \ref{fig:accessCountsWeek} shows weekly and daily patterns of
accesses. Similar to Figure \ref{fig:blogPostWeek}, we can see
periodic patterns of about 7 days, though the shape is not as smooth
as that in Figure \ref{fig:blogPostHour}.
Figure \ref{fig:accessCountsDay} shows the number of clicks on other
blogger's posts for each hour in a day.
The overall trend is more uniform compared to Figure
\ref{fig:blogPostHour}. 
Interestingly, this observation is consistent with ``cut and paste''
users in \cite{analyzingPatternsOfUser}. This suggests less serious
activities on the web, such as browsing or ``cut and paste'' editing,
follow a more uniform pattern than that for more serious activities.



%


\section{Influence Definition}\label{sec:influence}
As we have mentioned in Section \ref{sec:introduction}, there are two
components in the definition of social influence---(1) thoughts or
actions and (2) the thoughts or actions should be a result of being
affected by others.  In this section, we investigate these two
components, namely \textit{action} and \textit{causality}, in detail
and propose a framework for identifying influence in the blogoshere
with a high confidence level.

\subsection{Action Selection}
In our blog data set, the most frequent action is that a blogger (say
$A$) clicks on a post (say $p$), which was written by another blogger
(say $B$).  However, such an action by itself is not very helpful in
identifying influence among bloggers.  This is because the action of
clicking on $p$ usually takes place \textit{before} $A$ knows about
the content of $p$.  $A$ may have learned about $p$ from the front-page
of the web site, or she may be a loyal reader of $B$ and therefore
follows every post written by $B$.  In either case, the action of
clicking on $p$ is not necessarily influenced by the content of $p$.

\begin{figure}
\centering
\includegraphics[width=\linewidth]{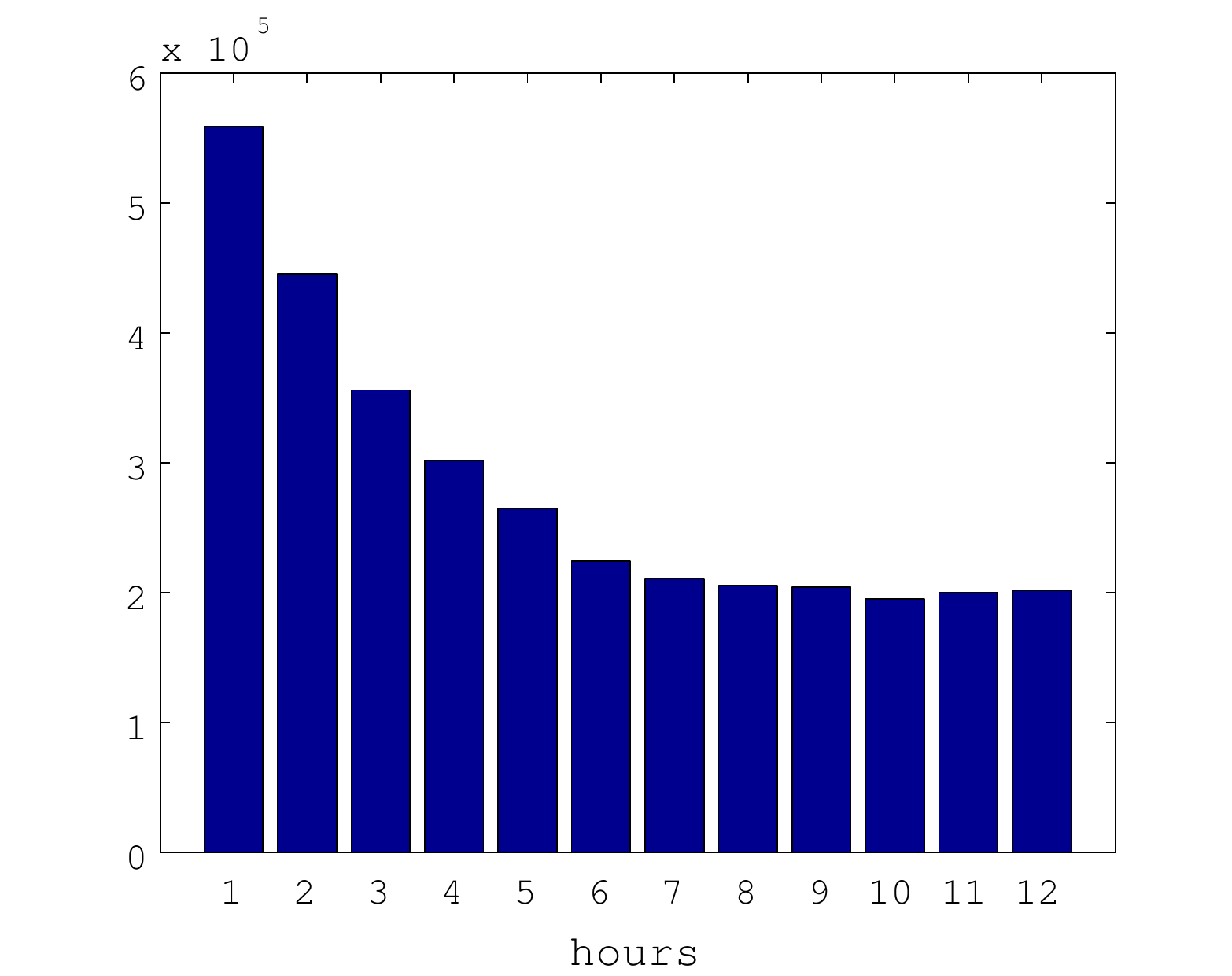} 
\includegraphics[width=\linewidth]{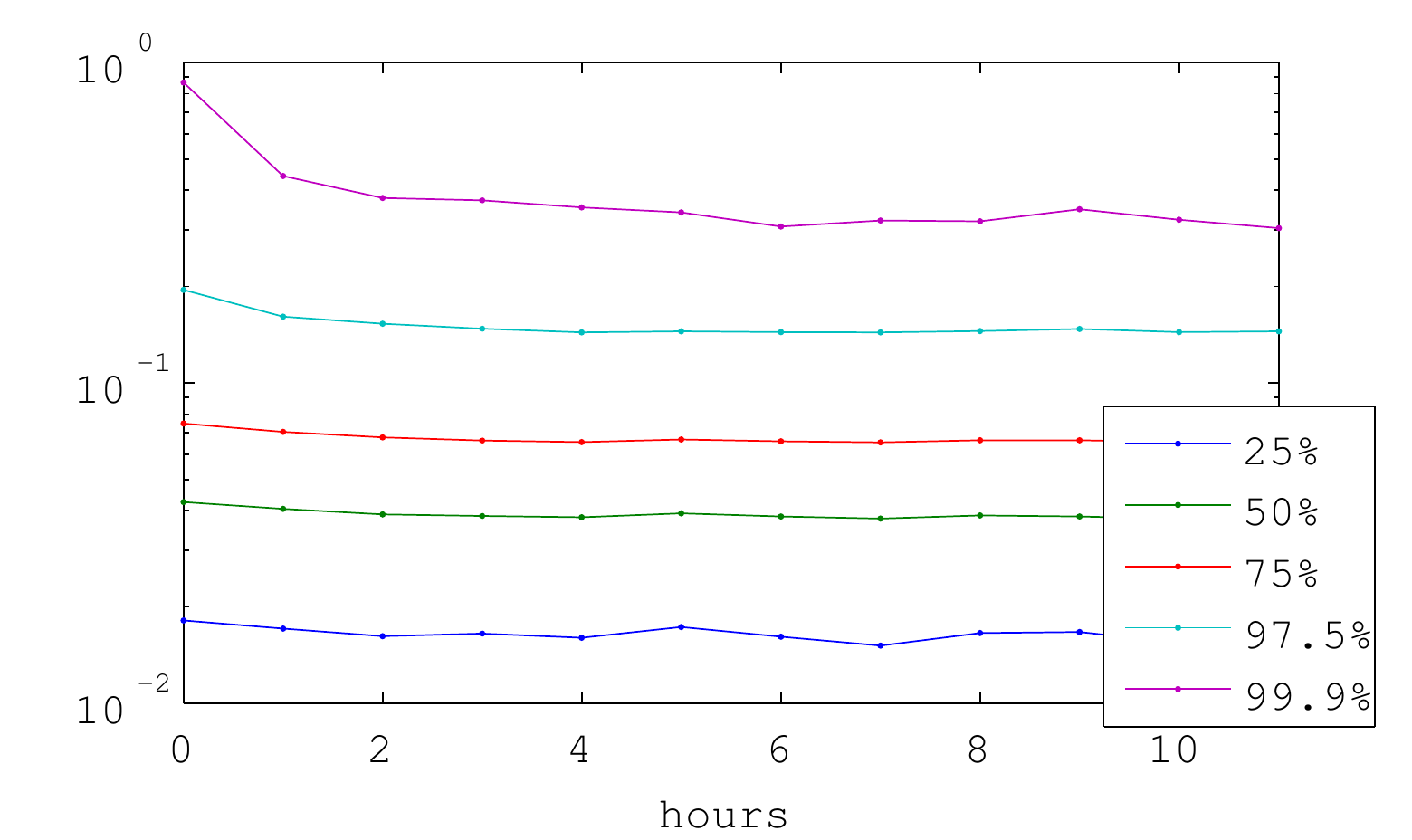}\\
{\hspace{-2cm}\scriptsize{ (a) \hspace{3.7cm} (b)}}
\caption{(a) Volume and (b) quantile plots of implicit link 12 hours
 before uploading. Each point in (b) corresponding one bin in (a).} \label{fig:implicitVolume}
\end{figure}

So instead, we focus on the action that a blogger writes a post.
Because the content of a post presumably reflects its author's
thoughts at the moment of writing, in the following discussion, we
consider interchangeably the content of a post and the thoughts of its
author at the time when the post is uploaded.  While the thoughts of a
blogger can be affected by many factors, such as some news she just
learned about from online news sites or from TV, because we want to
investigate the influence \textit{among bloggers}, we restrict the
factor to be the posts she has read before she writes her own post.
For this purpose, we build a post-level network consisting of
\textit{implicit links} in the following way.  We say there is an
implicit link from $q$ (written by $A$) to $p$ (written by $B$) if $A$
clicks on $p$ before she writes $q$.  That is, an implicit link
$(q,p)$ represents a high possibility that $q$ is influenced by $p$.
However, $A$ may have read many posts days or months before she writes
$q$ and it is not likely all of these posts have influence on $q$.  So
to reduce noises, we adopt a time window to remove $(q,p)$ pairs where
there is too large a time gap between when $A$ reads $p$ and when $A$
writes $q$.  To select a reasonable time window, we split the time
line into hourly buckets and in Figure~\ref{fig:implicitVolume}(a) we
plot the number of implicit links whose time gaps fall in each bucket.
(To avoid the spillover to the previous day, we limit the gap to be
less than 12 hours.)  As can be seen, the majority of implicit links
have time gaps of fewer than 5 or 6 hours.  For the purpose of
studying influence reflected by these implicit links, we set our time
window to be 12 hours.


The access logs used in building implicit links are cleaned in the
same way as that described in Section \ref{sec:accesslog}. The
statistics of the resulting network of implicit link is given as:

\begin{itemize}
\item Blogger: \\
  \#unique bloggers = 55,118

\item Post: \\
  \#unique posts = 1,268,979

\item Post level links (links between posts): \\
  \#unique implicit links = 3,847,172

\item Blogger level links (links between bloggers): \\
  \#unique implicit links = 877,540
\end{itemize}

\subsection{Causality Detection}
The implicit link we just defined captures influence among bloggers to
some degree---if we assume $A$ reads each post she clicked on and each
post she read influences her somewhat, both of which are reasonable
assumptions, then we can infer that all the implicit links from $q$
reflect certain influence on the content of $q$.  However, if we
simply treat implicit links as influence, we take a risk that the
confidence level for the resulting influence is too low.  The reason
for this is again causality vs. correlation, as we will describe next.

\subsubsection{Causality vs. Correlation}
For each implicit link $(q,p)$, there can be two possible explanations
for its occurrence.  The first explanation is that $A$ reads $p$
(written by $B$), gets influenced, and as a result, she decides to
write $q$.  This first explanation is described by the Bayesian
network in Figure~\ref{fig:t2}(a), which indicates a \textit{causal}
relation between $p$ and $q$.  The second explanation is that $A$
\textit{happens to} read $p$ and write $q$ whereas the former takes
place before the latter \textit{by chance}.  This explanation is
described by the Bayesian network in Figure~\ref{fig:t2}(b), which
indicates a \textit{correlated} relation between $p$ and $q$.  The
node denoted by \textbf{?} in Figure~\ref{fig:t2}(b) can be some
unknown reason that affects both $A$ and $B$, which explains why $A$
reads $B$'s posts in the first place.

As a result, in order to improve our confidence level on the influence
derived from the implicit links, we need to add additional
restrictions.

\subsubsection{Further Criteria}
We use two additional criteria to further improve our confidence level
on whether an implicit link really reflects influence.  We call these
two criteria the \textit{time similarity criterion} and the
\textit{content similarity criterion} and we give the definitions as
follows.
\begin{description}
\item[time similarity] For an implicit link $(q,p)$ to indicates
  influence, $A$ must read $p$ \textit{shortly before} she writes $q$;
\item[content similarity] For an implicit link $(q,p)$ to indicates
  influence, $p$ must have contents \textit{similar to} that of $q$.
\end{description}

\begin{figure}
\centering
\includegraphics[width=\linewidth]{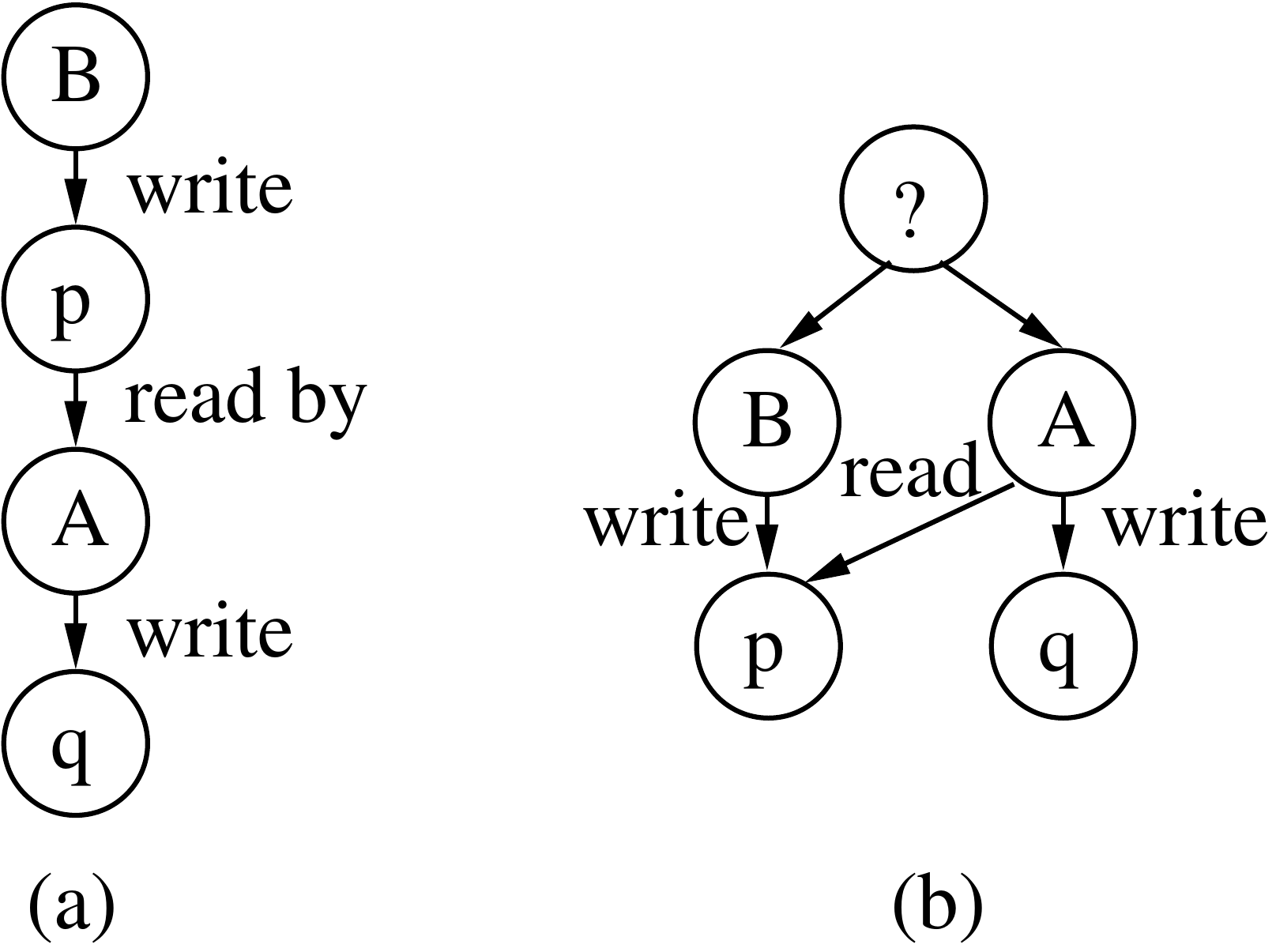} 
\caption{Two explanations for the implicit link $(q,p)$, where (a)
  indicates a causal relationship and (b) indicates a correlated
  relationship.} \label{fig:t2}
\end{figure}

The intuition behind the first criterion is that $q$ is more likely to
be affected by $p$ if $A$ reads $p$ \textit{shortly before} she writes
$q$.  This intuition is supported by psychology where it is well known
that short-term memory and attention span for human beings are both
rather short .  The intuition behind the second criterion is that for
$p$ to influence $q$, the content of $p$ should be somewhat
\textit{similar to} that of $q$.  For example, $A$'s thoughts on
healthcare reform, which she puts in $q$, is unlikely to be influenced
by $p$ if the content of $p$ is about a sports game.  And this is true
even $A$ reads $p$ minutes before she writes $q$.

To further verify these intuitions, we conduct the following
experiment to examine the content similarity between $q$ and $p$ for
all the implicit links $(q,p)$'s.  First, all the posts are put into
the bag-of-word representation and transformed into word vectors by
using a morphological analysis engine, where in the word vectors, only
nouns are kept (the corpus is all in Japanese).  Next, for each post
$q$ written by $A$, we locate all the posts $\{p_1,\dots,p_N\}$ read
by $A$ within 12 hours before $q$ is written.  Then for
$\{p_1,\dots,p_N\}$, we compute their similarities $\{c_1,\dots,c_N\}$
to $q$, i.e., $c_i = sim(q,p_i)$.  For the similarity, we choose the
cosine between $q$ and $p_i$.  Note that to ensure the cosine values
are reliable, we only compute for those implicit links $(q,p)$ where
the numbers of tokens (nouns) for $q$ and $p$ are both at least 10.
Finally, we assign to $c_i$ the time gap (in hours) on the implicit
link $(q,p_i)$.  Figure~\ref{fig:implicitVolume}(b) shows the quantile
plots of cosine similarity values (of the implicit links) with time
gap of 1 to 12 hours.  The figure clearly shows increased similarity
among implicit links with shorter time gaps.  This trend is
reassuring, because it shows the time similarity criterion and the
content similarity criterion tend to be consistent and together they
are likely to give us higher confidence level about if an implicit
link actually reflects influence.

However, both the two criteria still have their problems. For the
content similarity criterion, because an \textit{absolute} similarity
is used, it still suffers from the impact of correlation.  To
illustrate this point, we again look at the Bayesian network in
Figure~\ref{fig:t2}(b).  Assuming the node denoted by \textbf{?} is
something that highly impacts the contents of $p$ and $q$ (e.g., it
may represent some interests shared by $A$ and $B$), then $p$ and $q$
tend to be similar whether $q$ is influenced by $p$ or not.  Taking a
more extreme example, if $A$ reads and writes \textit{only} about the
healthcare reform and nothing else, then all the implicit links from
$A$ have high cosine similarities because of the narrow scope; in such
a case, we will draw an incorrect conclusion that $A$ is more likely
to be influenced.  The flaw of the time similarity criterion is that
there lacks a rigorous way to tell how shortly is good enough to have
causality dominating correlation.  To solve these problems, we fixed
the time similarity threshold and use a post-level \textit{relative}
similarity in our criteria and revise them as:
\begin{description}
\item[time similarity (revised)] For an implicit link $(q,p)$ to indicates
  influence, $A$ must read $p$ within $\tau$ hours before she writes
  $q$;
\item[content similarity (revised)] For an implicit link $(q,p)$ to indicates
  influence, the content of $p$ must be more similar to that of $q$
  than an \textit{average} implicit link from $q$.
\end{description}
We explain the two revised criteria in detail in the following.

\subsubsection{Time Shuffle Test}
To clarify the above two revised criteria and to make them concrete,
we first describe a time shuffle test that we designed to distinguish
the similarity due to influence and that due to correlation.  

\begin{figure}[htb]
\centering
  \includegraphics[width=\linewidth]{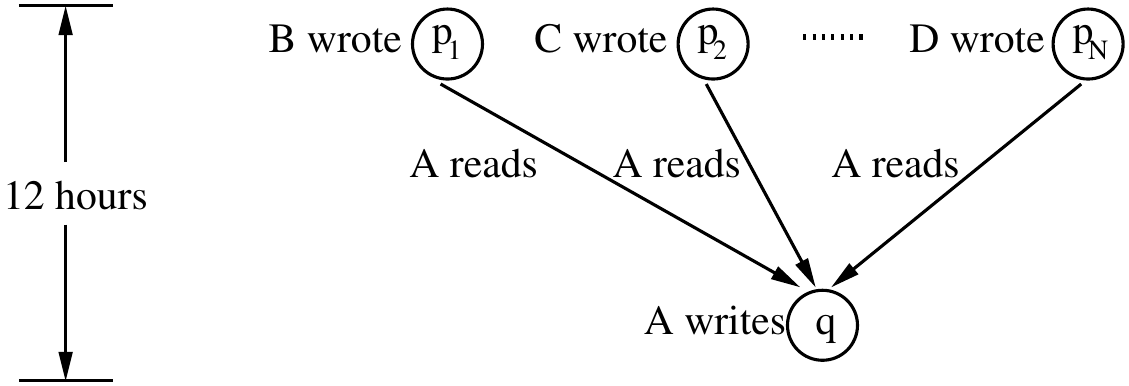} 
\caption{Illustration of the $z$-test.} \label{fig:z-illustration}
\end{figure}

The basic idea of the test is based on the assumption that correlation
is time-invariant (at least it is relatively homogeneous within the 12
hour time window in our data) and therefore should be insensitive to a
shuffle on the time line.  More specifically, for each post $q$, we
define a fair coin in the following way.  First we locate all the
posts $\{p_1,\dots,p_N\}$ and compute the similarity
$\{c_1,\dots,c_N\}$ in the same way as described before.  After that,
we find the \textit{median} among $\{c_1,\dots,c_N\}$, which we refer
to as $c_k$.  Then we turn $c_i$'s into heads (H) or tails (T)
according to whether $c_i$ is greater or less than the median $c_k$
(ties are broken randomly) to get something like
$\{H_1,T_2,\dots,H_N\}$.  Following that, according to the time gap on
the implicit link $(q,p_i)$, we put $H_i$ (or $T_i$) into the
corresponding hour bucket.  And we repeat these steps for each post,
to get a series of hour buckets with certain numbers of heads and
tails inside each bucket.  The experimental setting of this time
shuffle test is given in Figure~\ref{fig:z-illustration}.

Notice that in such a fair-coin design, instead of an
\textit{absolute} metric values, the \textit{relative} values with
respect to \emph{the median} are used and as a result, the exact
similarity metric used (cosine or Euclidean distance) is less
relevant.  In addition, because a fair coin is used for \textit{each
  post}, the similarity bias due to the shared node, denoted by
\textbf{?} in Figure~\ref{fig:t2}, is eliminated.  To see this point,
we go back to the previous extreme example: even if $A$ only read
posts about healthcare reform before she writes $q$, these posts are
still different in terms of how similar they are to $q$. So we still
can rank them with respect to $q$ and obtain a fair coin.  After the
fair coin is obtained, the exact similarities (which may be skewed, as
we discussed) are not relevant anymore.

After the buckets are collected, for each bucket we conduct a
one-sample $z$-test.  In this $z$-test, the statistics is the number
of heads and tails in each bucket.  The null hypothesis is that the
bucket is generated due to correlation (therefore fair coins) and the
alternative hypothesis is that the coins in the bucket are not fair.
Notice that the fair-coin null hypothesis is equivalent to a
fictitious test where all the $p_i$'s are shuffled randomly in the
time line, which should give fair coins in each hour bucket.  The $z$
value for this hypothesis test is $ z=(\bar{X} -
\mu)/(\sigma/\sqrt{n}), $ where $\mu=0.5$ (for the null hypothesis),
$\bar{X}$ is the sample mean (i.e., the fraction of heads in the
bucket), $\sigma^2$ is the variance, and $n$ is the number of samples
(i.e., the number of coins in the bucket).

\begin{figure}[ht]
\centering
  \includegraphics[width=\linewidth]{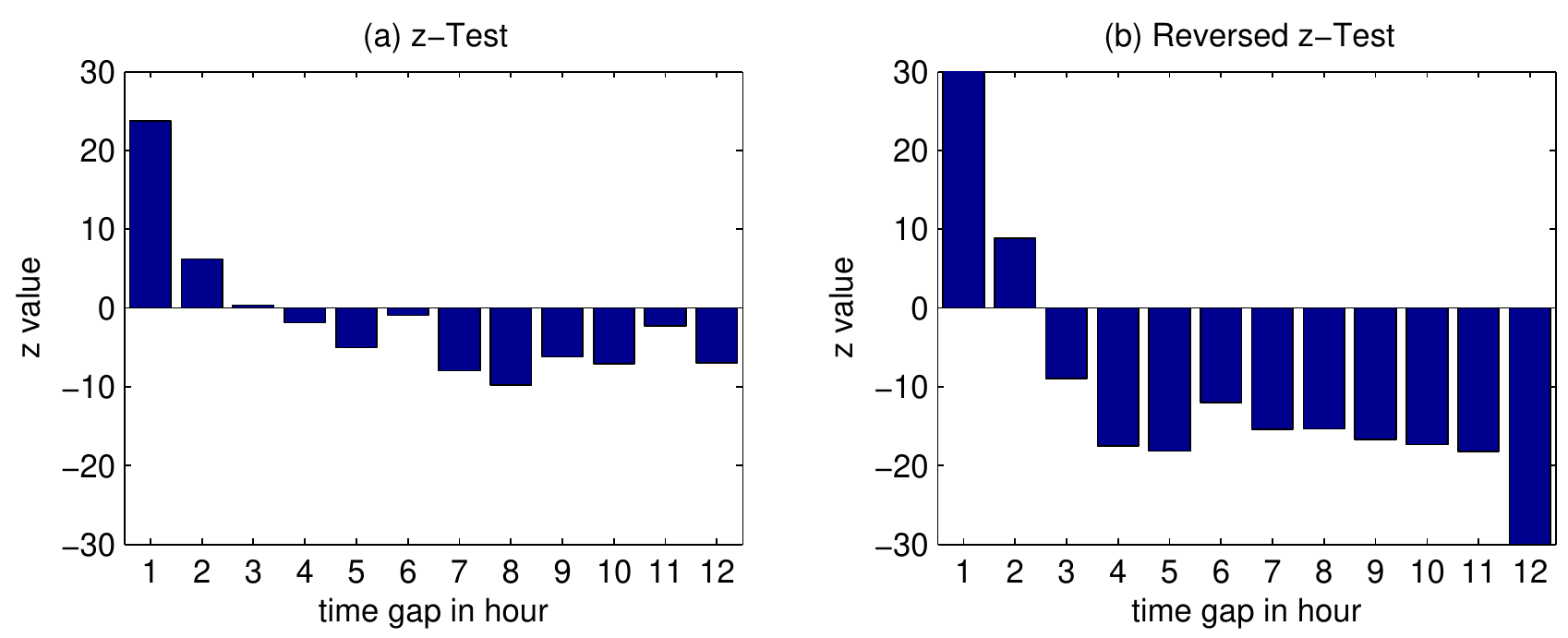} 
\caption{Results of the hypothesis test: (a) one sample $z$-test, (b)reversed $z$-test.} \label{fig:z}
\end{figure}

Figure~\ref{fig:z}(a) gives the $z$ values for the one sample tests
for the 12 hour buckets.  It is obvious that the number of heads in
each bucket should follow a Binomial distribution.  However, because
of the large number of coins in each bucket (usually tens of
thousands), we can approximate the distribution accurately by a normal
distribution.  Under such an approximation, if we choose a $p$-value
of 0.01, which is very typical in statistical analysis, we can reject
the null hypothesis in the buckets in hours 1 and 2.  It is
interesting to notice that we can also reject the null hypothesis in
some later hours, but with the conclusion that they are \textit{less}
similar to $q$ than what could be explained by correlation.  This
effect is due to the way the coins are designed---the total number of
heads is fixed (to be exactly half of all the coins) over the 12
buckets and so if the first two buckets contain more heads, the rest
ones will contain less.

\subsubsection{Reversed Time Shuffle Test}
\begin{figure}[ht]
\centering
  \includegraphics[width=\linewidth]{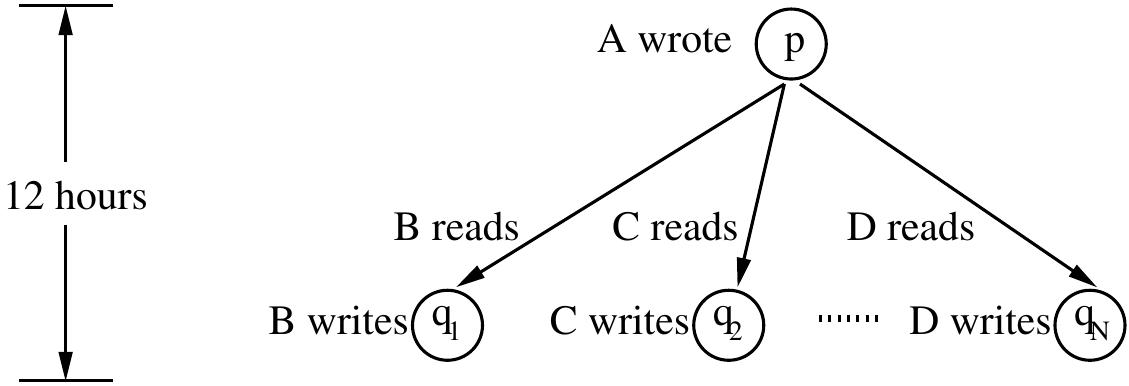} 
  \caption{Illustration of the reversed
    $z$-test.} \label{fig:z-reverse-illustration}
\end{figure}

In addition to the $z$-test just described, we also conduct a reversed
$z$-test in the following way.  We first reverse the direction of the
implicit links by changing $(q,p)$ to $(p,q)$ and keeping the original
time gap on the links.  Then we conduct the same $z$-test on the
reversed network of implicit link.  In other words, this time we
define a fair coin for each $p$ instead of for each $q$. And we want
to see among all the posts $q_j$'s that are written shortly after $p$
was read, if the similarity between $q_j$'s and $p$ are noticeably
different for different time gaps.  The experimental setting of this
reversed time shuffle test is given in
Figure~\ref{fig:z-reverse-illustration}.

Figure~\ref{fig:z}(b) shows the result of the reversed $z$-test.  As
can be seen, again the null hypothesis is rejected for the first two
hours, except that the $z$ values are much significant.  

A possible explanation for the more significant result of the reversed
$z$-test (in comparison to that of the $z$-test) is that for the
$z$-test, the coins are defined among $\{p_1,\dots,p_N\}$, which are
what $A$ read within 12 hours, and so they are more uniform (assuming
$A$'s interests do not change dramatically within 12 hours).  In
comparison, in the reversed $z$-test, the coins are defined among
$\{q_1,\dots,q_M\}$, which are written by different bloggers over
possibly much different time, and therefore they tend to be more
diversified.

As a result of the time shuffle test and that of the reversed time
shuffle test, we set the time $\tau$ in the time similarity criterion
to be 2 hours.

\subsection{Put It All Together}
With the identification of the appropriate actions and the criteria
for improving confident level, we finalize our definition of influence
as the following:

\begin{quote}
  We say that post $q$ (written by blogger $A$) is influenced by post
  $p$ if (1) $q$ is written within 2 hours after $p$ is read by $A$
  and (2) $p$ is more similar to $q$ than the similarity median among
  all posts read by $A$ within 12 hours before $q$ is written.
\end{quote}

With such a definition of influence, we are able to build, with
reasonable confidence level, the influence network among posts and
therefore among bloggers in our data set.  The statistics of the
resulting influence network is given as:

\begin{itemize}
\item Blogger: \\
  \#unique bloggers = 12,790

\item Post: \\
  \#unique posts = 717,304

\item Post level links (links between posts): \\
  \#unique implicit links = 487,282 

\item Blogger level links (links between bloggers): \\
  \#unique implicit links = 140,383
\end{itemize}

\subsection{Further Experimental Verifications}

\begin{figure}[htbp]
  \centering
\includegraphics[width=\linewidth]{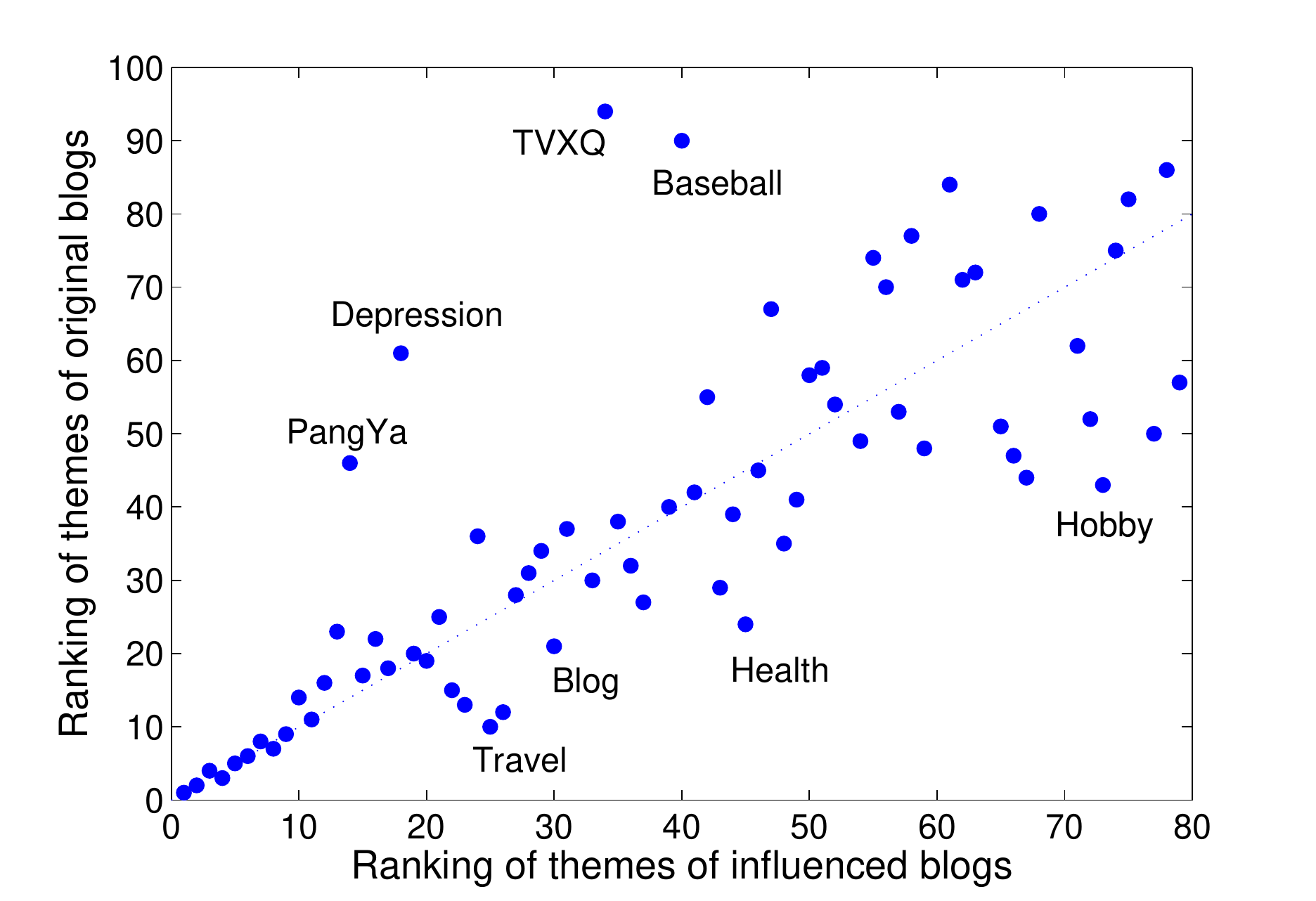}\\
\caption{Rankings of themes among original vs. influenced posts, where
  themes on the diagonal line have the same ranks in both the
  cases.}\label{fig:rank2D}
\end{figure}

As a sanity check of the above influence definition, we compare the
top themes among all the posts and those among the posts in the
influence network.  Figure~\ref{fig:rank2D} shows the scatter plot of
these top ranked themes.  A theme in the upper-left corner of the
figure indicates the rank of the corresponding theme is
\textit{promoted} in the influence network; a theme in the lower-right
corner indicates its rank is \textit{demoted} in the influence
network.  In the figure, we also show the name of several themes whose
ranks dramatically changed (either promoted or demoted).  As can be
seen, the themes got demoted the most in the influence network are
mainly about solo activities (travel, health, hobby) and those got
promoted the most are mainly about group activities (baseball,
PanYa---a multi-player online game, and TVXQ---about celebrity
gossip).  Another interesting observation we can get from the figure
is that \textit{depression seems to be rather contagious}!

\begin{figure}[htb]
  \centering
\includegraphics[width=\linewidth]{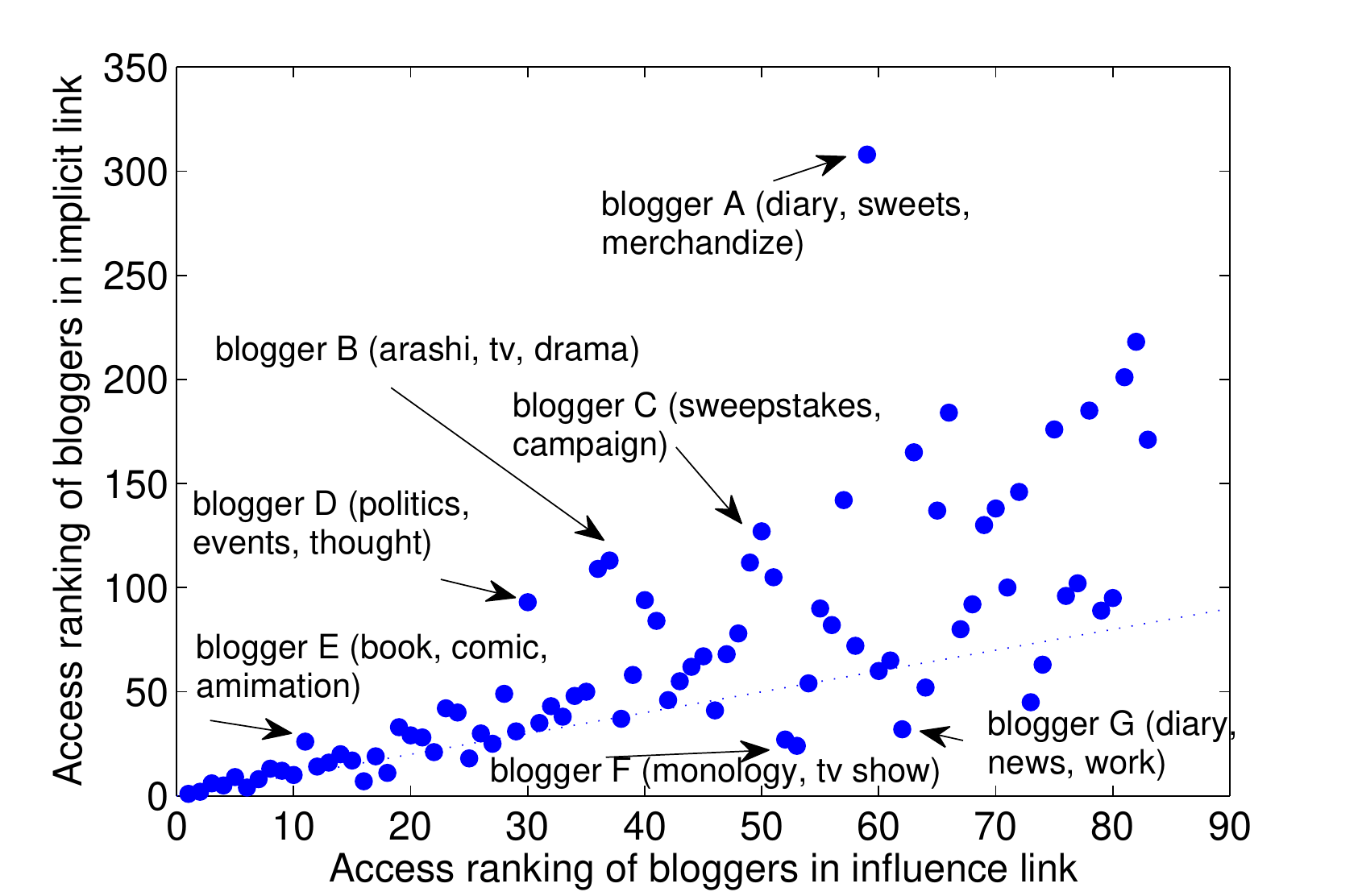}\\
\caption{Access rankings of bloggers among original vs. influenced
  bloggers, where bloggers on the diagonal line have the same ranks in
  both the cases. Blogger ID's are
  anonymized.}\label{fig:accessRank2D}
\end{figure}

Furthermore, to see how the popularity of bloggers changes in the
influence network, we show the scatter plot of access rankings of
influence network and implicit link network in
Figure~\ref{fig:accessRank2D}. Just like the theme ranking, bloggers
in the upper-left corner are promoted and those in the lower-right
corner are demoted in the influence network.  Clearly, we can see many
bloggers placed above the {\it diagonal} line, with greater lifts in
the ranks compared to Figure~\ref{fig:rank2D}. Note the scale
differences in the vertical axes. This result suggests that many
bloggers who are not the most popular ones, judging by how often they
are accessed, are actually very influential, judging by how often they
are thought and action provoking. In the figure, we annotate some {\it
  anonymized} representative bloggers with their frequent themes (in
parenthesis). As can be seen, topics such as sweets, arashi (Japanese
idol group), politics, and animation are indicative of existence of
{\it active} communities of bloggers influenced by on such
themes. Note with our influence definition, such communities are
formed not simply because of common interests, but because of actively
participating discussions and influencing each other.  On the other
hand, topics such as diary, monologue are indicative of solo
activities, the same as they were in the theme ranking.


\section{Influence on Different Topics}\label{sec:topic}
After the influence is successfully extracted, in this section and
the next, we analyze the influence in the Webryblog data by applying
several analytical algorithms that we recently developed
\cite{yun09:_iolap,combiningLinkAndContentForCommunityDetection}.
We mainly seek answers to two questions: if influence varies over
different topics and if influence varies over different members.  We
study the first question in this section and the second question in
the next section.

\subsection{Influence and Topics}
\textit{Are there different influential bloggers on different topics?}
Intuitively, the answer should be \textit{yes}.  In social science,
there are several well recognized factors that determine influence,
including charisma, reputation, bully pulpit, and peer
pressure.\footnote{\texttt{http://en.wikipedia.org/wiki/Social\_influence}}
Among these factors, arguably the most important one for the
blogosphere is \textit{reputation}.  This is because in the
blogosphere, most bloggers are ordinary people (hence charisma and
bully pulpit are less important) and not necessarily know each other
in the physical world (hence peer pressure is less significant).  As a
result, blogger $A$ reads blogger $B$'s posts probably mainly due to
$B$'s reputation on a given topic.  For $B$ to be a top influential
blogger, i.e., having many readers and having these readers frequently
posted responding posts, $B$ must have perceived expertise and
credibility.  It is not likely for a blogger to have such expertise
and credibility in \textit{all} topics.  This probably is the reason
why for each of the top-100 most popular (influential) bloggers listed
at Technorati, we can almost always assign a unique tag (such as
politics, technology, celebrity gossip, etc.) to the blogger, which
implies that a blogger is usually influential only on one topic.

Such an intuition sounds reasonable.  However, it is a challenging
problem to \textit{quantitatively} analyze the diversity of influence
of different topics .  For this purpose, we analyze the influence
network obtained in the previous section by using two state-of-the-art
analytical algorithms.  The two
algorithms---iOLAP~\cite{yun09:_iolap} and
PCL-DC~\cite{combiningLinkAndContentForCommunityDetection}---are
recently developed by us for social network analysis.  In addition, we
introduce a novel metric to quantify the diversity of influence over
different topics.  We first give a brief overview of these two
algorithms.

\subsection{Analytical Algorithms}

\subsubsection{iOLAP---an Approach based on Non-negative Tensor Factorization}

The first algorithm, iOLAP uses non-negative tensor factorization to
analyze polyadic data (those data with higher dimensions than
traditional dyadic, or matrix data).  For this Webryblog data set, we
build the polyadic data in the following way.  Assume we have
determined that $p$ (written by $B$) has influence on $q$ (written by
$A$), then for each keyword $w$ shared by $p$ and $q$ (such a shared
keyword always exists because the cosine similarity between $p$ and
$q$ is nonzero according to our definition of influence), we generate
a triple $\langle A,B,w \rangle$.  Such a triple is a piece of
evidence that \textit{$A$ is influenced by $B$ on $w$}.  By collecting
all such triples that can be derived from the influence among all
bloggers, we obtain a tensor $\calD$ of dimension $b\times b\times v$,
where $b$ is the number of bloggers and $v$ is the size of the
vocabulary.  $\calD_{ijk}$ is the frequency that blogger $i$ is
influenced by blogger $j$ on keyword $k$.  After $\calD$ is
constructed, we apply on $\calD$ the iOLAP algorithm, which seeks the
optimal parameters that maximize the data log-likelihood, i.e.,
\[
\argmax_{\Theta} \sum_{i,j,k} \calD_{ijk} \log \left(\sum_{i'j'k'}\calC_{i'j'k'}X_{ii'}Y_{jj'}Z_{kk'}\right)
\]
where the parameter $\Theta$ consists of $X,Y,Z$, and $\calC$; $X\in
R_{b\times I}, Y\in R_{b\times J}$ and $Z\in R_{v\times K}$ are the
major components of influencing bloggers, influenced bloggers, and key
topics, respectively; and $\calC \in R_{I\times J\times K}$ is a core
tensor (of dimension $I\times J\times K$, which is much smaller than
the dimension $b\times b\times v$ of the original data $\calD$) that
captures the interaction among the major components in $X$, $Y$, and
$Z$.  The high-level idea of the iOLAP algorithm is that, the
parameter $\Theta=\{\calC,X,Y,Z\}$ learned by the iOLAP algorithm
captures $I$ most significant groups of influential bloggers, $J$ most
significant groups of influenced bloggers, $K$ most significant sets
of topics, and the relationship among them.  From the learned
parameters, we are able to derive the top influential bloggers on each
topic.


\subsubsection{PCL-DC---an Approach based on Stochastic Block Model}

The second algorithm, PCL-DC, is an improvement over the well known
stochastic block model.  It uses a conditional link model for link
analysis and a discriminative approach to model the content.  PCL-DC
learns the parameters to maximize the data log-likelihood as
\[
\argmax_{\Theta}\sum_{(i\rightarrow j)\in\mathcal E}
s_{ij}\log\sum_ky_{ik}\frac{y_{jk}b_j}{\sum_{j'\in\mathcal{LO}(i)}y_{j'k}b_{j'}}
\]
where $y_{ik}=\exp(w_k^Tx_i)/\sum_l \exp(w_l^Tx_i)$, is the logistic
discriminative model on contents.  In the formula, $b_i$ indicates the
popularity (influence) of blogger $i$, $s_{ij}$ is the number of links
from $i$ to $j$, $\mathcal{E}$ represents the set of links,
$\mathcal{LO}(i)$ represents the set of all bloggers that have
influence on $i$, and $x_i$ is the topic vector for blogger $i$.
$\vec{b}\in R_b$, $Y\in R_{b\times K}$ and $W\in R_{v\times K}$ are
the parameters to learn.  The high-level idea of the PCL-DC algorithm
is that, the parameter $\Theta=\{\vec{b},Y,W\}$ of the PCL-DC
algorithm captures following: the influence of each blogger, for each
blogger how her influence is distributed among different topics, and
the contents of the topics, respectively.

In addition, for the purpose of this paper, we revise the iOLAP and
PCL-DC algorithms so that they share the same predefined set of
topics.  The main reason for this restriction is to make the results
of the two algorithms comparable.  The predefined $K$ topics (the
number $K$ is, rather arbitrarily, set to 50) are obtained by using
the well known PLSA algorithm.  Table~\ref{tab:topicKeyword} shows the
top keywords in several representative topics. From the keywords we
can see that the topics are both unambiguous and well separated.


\begin{table}[ht]%
\tbl{Top keywords (translated from Japanese) in some representative
 topics.\label{tab:topicKeyword}}{%
\begin{tabular}{|l|p{0.5\hsize}|}
\hline \hline
T1 &\baselineskip=5pt
{\footnotesize recipe, taste, salt, vegetables, ingredient, water, cake,
 rice, meal, meat, salad, sugar, bread, lunch}\\ \hline
T2& \baselineskip=5pt
{\footnotesize family, genus, leaf, flower, color, plant, seed, garden,
     stem, grass, name, shape, spring, autumn} \\ \hline
T3& \baselineskip=5pt
{\footnotesize rakuten market, shipping fee, goods, store, size,
     purchase, sale, price, free, item, popular} \\ \hline
T4& \baselineskip=5pt
{\footnotesize love, heart, hand, human, feel, meaning, life, friends,
     hand, world, force, image, boy friend} \\ \hline
T5& \baselineskip=5pt
{\footnotesize runs, starter, pitcher, game, hit, baseball, loss
     allowed, clutch, three strikes, batting order} \\ \hline
T6& \baselineskip=5pt
{\footnotesize hospital, remedy, examination, medical, doctor, pain,
     symptom, exercise, weight, diet, result} \\ \hline
T7& \baselineskip=5pt
{\footnotesize offense, gundam, enemy, harness, damage, game, recover,
     weapon, level, state, clear, point} \\ \hline
T8& \baselineskip=5pt
{\footnotesize morning, work, house, holiday, evening, dinner, lunch,
     shopping, yesterday, tomorrow, home } \\ \hline
T9& \baselineskip=5pt
{\footnotesize rail, train, station, travel, hotel, platform, express,
     bus, arrival, line, departure, sight seeing} \\ \hline
T10& \baselineskip=5pt
{\footnotesize mountain climbing, path, peaks, ridge, bifurcate,
     descent, departure, course, arrival, parking} \\ \hline\hline
\end{tabular}}
\end{table}%
\subsection{Metric for Influence Diversity}
To quantitatively measure the diversity of top influential bloggers
computed by iOLAP and PCL-DC on different topics, we introduce a novel
metric, which we termed the \textit{influence diversity ratio} (IDR).
Here is how IDR is computed.  Assume $K$ is the number of topics.  For
any given positive integer $N$, we can compute the total number $C_N$
of \textit{distinct bloggers} among the top-$N$ most influential
bloggers among \textit{all the topics}.  Intuitively, a lager $C_N$
indicates that for the fixed number of $K$ topics, more bloggers show
up in the top-$N$ most influential bloggers and therefore the top-$N$
influential bloggers are more diversified.  $C_N$ obviously ranges
between $N$ and $K N$.  By normalizing $C_N$, we define the
influence diversity ratio at $N$ as
\[
IDR_N = (C_N-N)/[N\cdot(K-1)].
\]
Notice that for any $N$ and $K$, $IDR_N$ is always between 0 and
1. $IDR_N=0$ when the \textit{same} set of $N$ bloggers are ranked the
top-$N$ most influential ones among all the $K$ topics.  On the other
hand, $IDR_N=1$ when there is no overlap between the top-$N$ most
influential bloggers of \textit{any two topics}.

\subsection{Results and Discussion}
\begin{figure}
\centering
  \includegraphics[width=0.8\hsize]{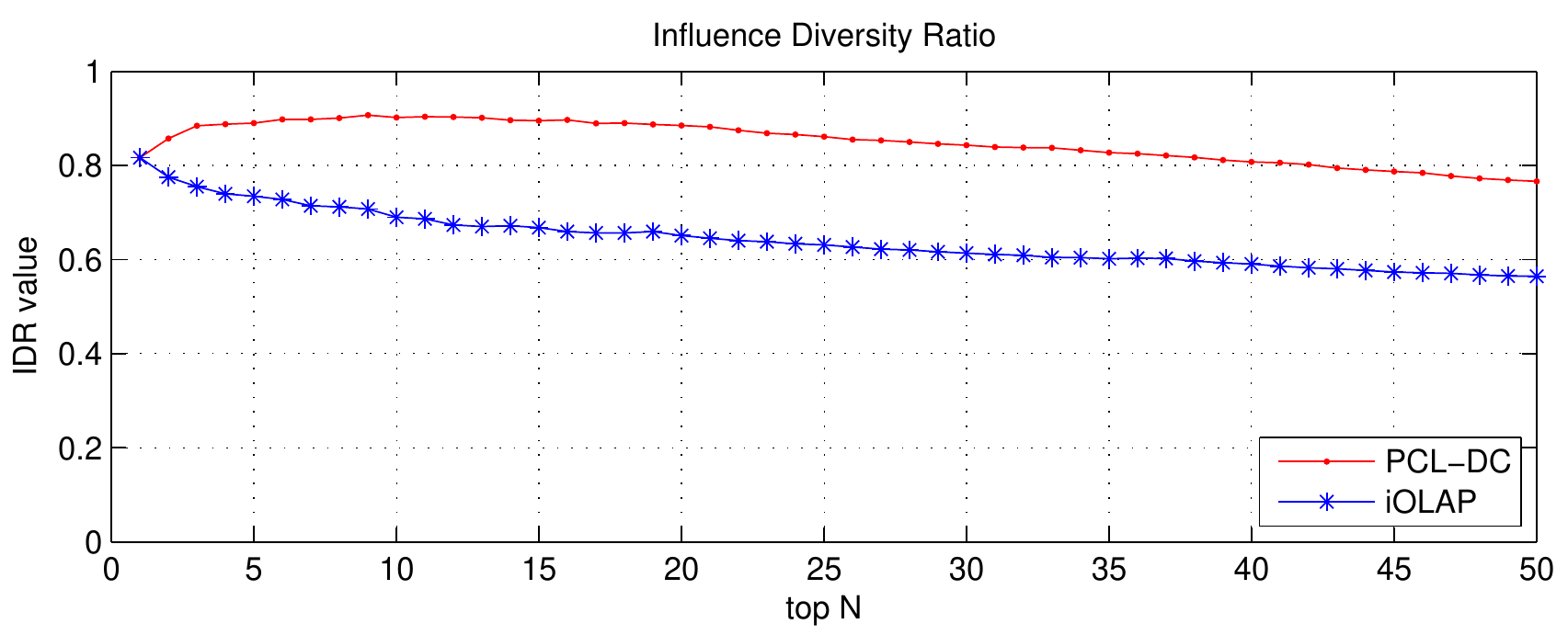} 
\caption{The influence diversity ratio at $N$ for iOLAP and PCL-DC results.} \label{fig:idr}
\end{figure}

Figure~\ref{fig:idr} shows the IDR$_N$ values for $N=$1 to 50 for the
most influential bloggers, derived by iOLAP and PCL-DC on the 50
topics. (Note that both the algorithms used the same set of 50
topics.)  From the figure we can observe the following.  First, the
$IDR_N$ values are rather high over different $N$ and are especially
high for smaller $N$ (for $N$ less than 10).  This result verifies our
intuition that the top influential bloggers among different topics
should be different.  Second, the $IDR_N$ values start to decrease
steadily as $N$ grows large (for $N$ greater than 20).  This result
suggests as $N$ increases, more and more top-$N$ influential bloggers
are shared among different topics.  Thirdly, comparing the $IDR_N$
values of PCL-DC with that of iOLAP, we can see than PCL-DC gives more
diversified top influential bloggers, especially for small $N$'s.
This difference may due to the nature of the two algorithms, where
iOLAP is a generative model and PCL-DC is a discriminative one.

Finally, we want to point out that we have conducted similar tests
under different topic numbers $K$.  The results, whose details are
skipped due to the space limit, turn out to be similar, which implies
that the exact number of topics is not a crucial factor.

\section{Influence on Different Members}\label{sec:experiment}
In the previous section, we investigated if influence is different on
different topics.  In this section, we ask the question ``if there are
different influential bloggers for different members, even on the same
topic''. Again, intuitively the answer should be \textit{yes}.  For
example, on the same topic of politics, a member with Democratic
leanings may be influenced by totally different people compared to a
member with Republican leanings.  That is, even on the same topic,
different members may have different beliefs or tastes and therefore
get influenced by different bloggers.  To study such a
\textit{personalized} influence, of course we can use techniques
similar to those used in the previous section.  Instead, however, to
verify that influence is personalized, we design an extrinsic test.  A
main reason for such a test, other than to verify influence in
personalized, is to show that the extracted influence can be directly
applied in practice to improve user experience in the blogosphere.

\subsection{Extrinsic Test}
For the extrinsic test, we use the task of personalized blogger
recommendation within the given influence network.  The problem is
described as: \textit{Given the historic activities of blogger $A$,
  and given a set of keywords $W$ that $A$ is interested in, can we
  recommend a blogger $B$, which $A$ has not read before, that will
  have high influence on $A$ on the give keywords $W$?}  One usage
scenario for this task is when a member asks the following query
\textit{``Show me top bloggers, among bloggers whose posts I have not
  read before, that will very likely to affect my thoughts on the
  issue of healthcare reform.''}  For such a recommendation, we can
directly use the topic-specific influential bloggers obtained in the
previous section.  That is, we recommend the most influential bloggers
on the keywords $W$ (we will show how to do it shortly).  However,
such a solution is a \textit{global} one in that each blogger gets
\textit{the same} ranked list of influential bloggers.  In contrast,
by using iOLAP and PCL-DC algorithms, we are able to make
\textit{personalized} recommendation by taking into consideration the
historic activities of the blogger to whom the recommendation is made.

If it turns out that the performance of the personalized
recommendation is considerably better than that of global
recommendation, then we can infer that personalization helps identify
influential bloggers for each member, which in turn indirectly proves
that influence is different for different members.

\subsection{Data Preparation and Algorithms}
For this recommendation task, because at the time of this writing we
still are not able to plug our recommendation engine into the
real-time BIGLOBE system, we simulate the recommendation task in the
following way.  We start by splitting the influence network into two
parts---a training set and a test set.  Starting from the influence
network, for each node $A$ in the network we locate all the bloggers
that have influence on $A$ (i.e., those nodes that are reachable from
node $A$ in one step in the influence network); then among these
bloggers, we randomly select one, say $B$, and remove link $(A,B)$
from the influence network and put $(A,B)$ into the test data set.
After applying this process to all the nodes, the remaining influence
network (with test data removed) is used as the training data set.
Notice that because of the way the data are split, for each $(A,B)$ in
the test data, $(A,B)$ is absent from the training data.  By doing
this, we avoid the ``bookmark effect'', where $A$ tends to read
bloggers that she has read before.  Furthermore, to make the influence
topic-specific, we provide to the algorithms the set of keywords $W$
on which $A$ is actually influenced by $B$.  That is, $W$ is the union
of overlapping keywords shared by any pair of $(q,p)$ in the test data
where $q$ is written by $A$ and $p$ is written by $B$.  In real
applications, $W$ can simply be the keywords (e.g., \textit{healthcare
  reform}) provided by the blogger in her query.

For the global recommendation, we use the topic-specific influential
bloggers in the following way.  If $B$ indicates a blogger, $W$
indicates the query keywords, and $t_k$ indicates the $k$-th topic,
then we can write the conditional probability (keyword-specific
influential blogger) of recommending blogger $B$ given keywords $W$ as
\[
P(B|W) = \sum_k P(B|t_k)\cdot P(t_k|W)
\]
where $P(B|t_k)$ is obtained in the previous section and $P(t_k|W)$
can be obtained from the output of iOLAP by using the Bayes rule:
$P(t_k|W)\propto P(W|t_k)\cdot P(t_k)$.  We refer this recommendation
as TG, for topic-specific global recommendation.

For the personalized recommendations, for iOLAP we have
\[
P(B|A,W) = P(A,B,W)/P(A,W) \propto [\calC,X,Y,Z]_{ABW}
\]
and for PCL-DC we can again use the Bayes rule.  

\subsection{Results and Discussion}
\begin{figure}
\centering
\includegraphics[width=0.8\hsize]{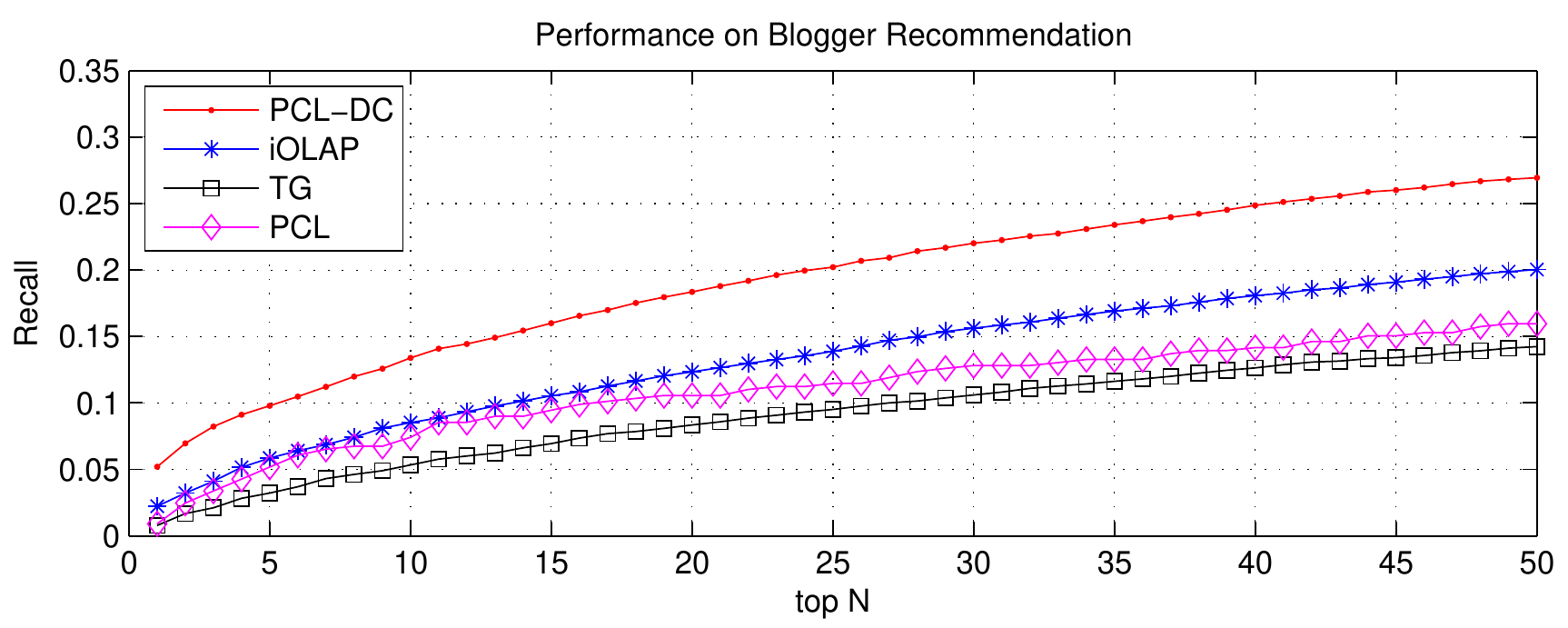}
\caption{Recall at top-$N$ for the blogger recommendation task.} \label{fig:recall}
\end{figure}

In terms of performance metric, we use a measure typically used in the
information retrieval field, \textit{recall-at-$N$}.  That is, if $N$
recommendations are allowed to be made to each member, what fraction
of the links in the test data are correctly recalled.
Figure~\ref{fig:recall} shows the performance for all the algorithms.
From the performance we can see that compared to the global
recommendation (TG), the personalized recommendations (iOLAP and
PCL-DC) have noticeably better performance than the global
recommendation, because they considered historic behaviors and
predicted the influential people for each member differently.  This
verifies our conjecture that influence differs over members, even on
the same topic.  In addition, we also show the performance of PCL, a
simplified version of PCL-DC but without content analysis.  The
relatively poor performance of PCL demonstrates that by using link
analysis alone we do not have an accurate model for the influence
network.  As a consequence, we have learned that a good model for the
influence network relies on both contents and links.  In other words,
influence is both topic-specific and member-specific.

\section{Conclusion and Future Work}\label{sec:conclusion}
In this paper we analyzed influence in a large blog data set.  We
defined influence in a principled way by selecting appropriate
actions, proposing intuitive criteria, and designing rigorous
statistical tests.  After the influence was extracted, we further
investigated the questions that if influence is topic-specific and if
it is member-specific.  We provided affirmative answers to these
questions by leveraging state-of-the-art algorithms for social network
analysis, introducing novel quantitative measures, and using both
intrinsic and extrinsic tests.  Some of the tests also reveals the
potential application of influence in blogger recommendation in the
blogosphere.  To our best knowledge, such an extensive analysis on
influence in such a large-scale data set is the first of its kind.

For future work, we plan to extend our investigation in the following
two directions. First, in this paper we essentially assumed a
time-invariant system where influence is considered static and topics
are treated as unchanged over time. This assumption may be
questionable when bloggers dynamically join and leave the system or
when the topics are time-sensitive (e.g., a president election). In
the future, we plan to study dynamics in this blog data set. Second,
in this paper we mainly focused on one-step direct influence. But
previous studies have observed cascade behavior in social networks
where information is diffused through paths with multiple
hops. Applying these diffusion models may lead additional insights
into our analysis and is one of our future directions.

\begin{acks}
The authors thank BIGLOBE for their support for preparing and providing us with the valuable data.
\end{acks}

\bibliographystyle{acmsmall}
\bibliography{referencesShort}



\end{document}